\def\year{2021}\relax
\documentclass[letterpaper]{article} 
\usepackage{aaai21}  
\usepackage{times}  
\usepackage{helvet} 
\usepackage{courier}  
\usepackage[hyphens]{url}  
\usepackage{graphicx} 
\usepackage{graphicx}
\usepackage{arydshln}
\usepackage{amsmath,amssymb,amsfonts}
\usepackage{algorithmic}
\usepackage{tabularx}
\usepackage{textcomp}
\usepackage{makecell}
\usepackage{xcolor}
\usepackage[switch]{lineno}
\usepackage{natbib}  
\usepackage{caption} 
\title{Embedding Heterogeneous Networks into Hyperbolic Space Without Meta-path
}

\author {
    Lili Wang,\textsuperscript{\rm 1}
    Chongyang Gao, \textsuperscript{\rm 1}
    Chenghan Huang,  \textsuperscript{\rm 2} 
    Ruibo Liu,  \textsuperscript{\rm 1} 
    Weicheng Ma,  \textsuperscript{\rm 1} \\
    Soroush Vosoughi   \textsuperscript{\rm 1} \\
}
\affiliations {
    \textsuperscript{\rm 1} Department of Computer Science, Dartmouth College \\
    \textsuperscript{\rm 2} Millennium Management LLC \\
    lili.wang.gr@dartmouth.edu, soroush.vosoughi@dartmouth.edu
}

\begin{document}

\maketitle

\begin{abstract}

Networks found in the real-world are numerous and varied. A common type of network is the heterogeneous network, where the nodes (and edges) can be of different types. Accordingly, there have been efforts at learning representations of these heterogeneous networks in low-dimensional space. However, most of the existing heterogeneous network embedding methods suffer from the following two drawbacks: (1) The target space is usually Euclidean. Conversely, many recent works have shown that complex networks may have hyperbolic latent anatomy, which is non-Euclidean. (2) These methods usually rely on meta-paths, which require domain-specific prior knowledge for meta-path selection. Additionally, different down-streaming tasks on the same network might require different meta-paths in order to generate task-specific embeddings. In this paper, we propose a novel self-guided random walk method that does not require meta-path for embedding heterogeneous networks into hyperbolic space. We conduct thorough experiments for the tasks of network reconstruction and link prediction on two public datasets, showing that our model outperforms a variety of well-known baselines across all tasks.
\end{abstract}

\section{Introduction}
There has been an ever-increasing amount of network data available to researchers in recent decades. Through encoding network structure into low-dimensional embeddings, researchers can mine and model networks without resorting to feature engineering.
 
In the field of natural language processing, there has been a long history of generating low dimensional embeddings for words based on co-occurrence in the same contexts, which is similar to the local proximity in a graph. Using language models to embed networks was first proposed by Perozzi et al. in their work DeepWalk \cite{deepwalk}. The random walks are adopted to generate the context of nodes from the network, which are then fed into a \emph{Skip-Gram} model \cite{word2vec} as sentences. This way, similar nodes in a network will have similar contexts and thus have related embeddings that are near one another. Grover et al. proposed an extension of DeepWalk called node2vec \cite{node2vec} by smoothly interpolating breadth-first and depth-first sampling, thus combining different views of node equivalence.

After these two pioneering papers, there has been an explosion of work focused on representation learning for different types of networks. One way to divide the prior work is to separate embedding methods for homogeneous and heterogeneous networks. Homogeneous networks contain only one type of node. Heterogeneous networks, on the other hand, have two or more types of nodes. For example, citation networks where all nodes are papers and all edges correspond to citations are homogeneous. However, if we extend the network to include authors and venues as nodes, then it becomes a heterogeneous network. In this network, the edges represent different kinds of relationships. For instance, author-to-author edges would represent collaboration, while author-to-paper edges would represent authorship and so on.

Various methods for embedding heterogeneous networks have been proposed (see for instance,  \citealt{metapath2vec,hin2vec,JUST,hetespaceywalk,Aspem,pte,HEER,HHNE}). However, to the best of our knowledge, all the proposed methods, with the exception of HHNE \cite{HHNE},  embed heterogeneous networks into Euclidean space. In this paper, we introduce a novel embedding method for heterogeneous networks in hyperbolic space.

Hyperbolic and Euclidean spaces can both preserve certain properties, such as angles between two vectors when the conformal transformation is used. However, there are two main advantages of hyperbolic space compared with Euclidean space. First, hyperbolic space has a stronger representation ability for hierarchical structures; for instance, \citeauthor{sarkar2011low} (\citeyear{sarkar2011low}) show that trees can be embedded into the two-dimensional hyperbolic space (Poincar\'e disk) with arbitrarily low distortion. On the other hand, Linial et al. (\citeyear{linial1995geometry}) show that Euclidean space cannot represent a tree with low distortion. A sizeable number of networks have a hierarchical structure, which makes them suitable for embedding into hyperbolic space. Another advantage of hyperbolic space is that it represents high-degree vertices better than Euclidean space, which is important when dealing with complex networks where the degree of the vertices follow a power-law distribution (i.e., where there are hubs with many large degrees).  

Having established the advantages of hyperbolic space for network embedding, we propose a novel method for embedding heterogeneous networks into hyperbolic space. As mentioned earlier, as far as we are aware the only other method proposed for embedding heterogeneous networks into hyperbolic space is HHNE \cite{HHNE}. However, HHNE suffers from the following disadvantages:
\begin{itemize}

    \item It requires meta-paths to guide random walk. Existing literature shows that the selection of meta-paths highly affects the quality of the learnt embeddings \cite{JUST}, and the selection of meta-paths needs domain-specific prior knowledge. This makes HHNE not easily generalizable.

    \item Even for the same network, HHNE requires different types of meta-path for training models for down-stream tasks of different types of edges (e.g. link prediction for paper-author and paper-venue edges). We believe it is preferable to have one general and consistent embedding for all down-stream tasks, reducing the need for training models multiple times.

    \item HHNE embeds networks to a \emph{Poincar\'e ball}. But to the best of our knowledge, the gradient on the \emph{Poincar\'e ball} cannot be calculated precisely and needs a first-order approximation to the exponential map, called a \emph{retraction} \cite{bonnabel2013stochastic}.
\end{itemize}

In this paper, we propose a novel self-guided random walk method that solves the issues listed above. Our method does not require meta-paths and uses very few non-sensitive parameters across all the domains and down-streaming tasks. We use the \emph{hyperboloid} model to embed networks in order to avoid the approximation of \emph{retraction}. We conduct thorough experiments for the tasks of network reconstruction and link prediction on two public datasets, showing that our model can achieve superior performance across all tasks.

\section{Related Work}

\subsection{Homogeneous Network Embedding}
  Besides random walk based methods DeepWalk and node2vec, LINE \cite{line} was proposed for large-scale network embedding, and can preserve first and second-order proximities. Cao et al. (\citeyear{grarep}), extended LINE by creating GraRep, which preserves k-step proximities when embedding networks. Their model captures the different k-step relational information with different values of k amongst vertices from the graph directly by manipulating different global transition matrices defined over the graph, without involving slow and complex sampling processes. SDNE \cite{SDNE} is also an extension of LINE which uses a semi-supervised deep autoencoder model to capture both first and second-order proximities in networks. HOPE \cite{HOPE} learns vertex representations that capture the
asymmetric high-order proximity in directed networks by solving a matrix factorization problem, while APP \cite{APP} is another network embedding method designed to capture asymmetric proximity by using a \emph{Monte Carlo} approach to approximate the asymmetric \emph{rooted PageRank} proximity \cite{root}. In the area of hyperbolic embedding, Nickel et al. (\citeyear{nickel2017poincare}) proposed a method (PoincaréEmb) of learning hierarchical representations of symbolic data; McDonald et al. (\citeyear{HEAT}) proposed a ``teleport" walk to embed attributed graphs and Ganea et al. (\citeyear{ganea2018hyperbolic}) use hyperbolic cones as a heuristic for embedding directed acyclic graphs. Our prior work \cite{wang2020embedding} also presents a method for embedding structural role similarity of nodes of homogeneous networks into hyperbolic space.

\subsection{Heterogeneous Network Embedding}

To deal with heterogeneous networks,
metapath2vec \cite{metapath2vec} extends DeepWalk and node2vec by a meta-path guided random walk. HIN2Vec proposes a multi-task learning method based on different relation types and network structures. JUST \cite{JUST} provides a Jump/Stay strategy of random walk to balance the heterogeneous edges and homogeneous edges. HeteSpaceyWalk \cite{hetespaceywalk} first formalizes the meta-path guided random walk to a high-order Markov chain process, and then utilizes an efficient random walk which they call heterogeneous personalized spacey random walk. AspEm \cite{Aspem} proposes the concept of aspects in networks, and preserves semantic information in heterogeneous information networks based on multiple aspects. PTE \cite{pte} decomposes a network to multiple bipartite networks by the type of edges to learn representations from the one-hop neighbourhood, and HEER \cite{HEER} extends it by considering typed closeness. In the area of hyperbolic embedding, there is only one heterogeneous embedding method, HHNE \cite{HHNE}, which extends metapath2vec by using meta-path guided random walk to embed nodes into a \emph{Poincar\'e ball}.
With the development of Graph Neural Networks (GNN), many method like R-GCN \cite{RGCN}, HGT \cite{hgt}, HAN \cite{HAN} and HetGNN \cite{HETGNN} utilize GNNs for heterogeneous networks. Finally, knowledge bases are a special kind of heterogeneous network. Many relation-learning based methods for knowledge bases like TransE \cite{transe}, TransH \cite{transh}, RotatE \cite{rotate}, DistMult \cite{dismult}, NKGE \cite{NKGE}, and SACN \cite{SACN} have been proposed.

\section{Framework}
\subsection{Preliminaries}
 A heterogeneous network is defined as a graph $G=(V, E, T)$, in which each node $v$ and each edge $e$ are associated with their mapping functions $\phi(v): V \rightarrow T_{V}$ and $\varphi(e): E \rightarrow T_{E}$ respectively. $T_{V}$ and $T_{E}$ denote the sets of object and relation types, where $\left|T_{V}\right|+\left|T_{E}\right|>2$. The task is to learn a d-dimensional hyperbolic embedding $\mathrm{X} \in \mathbb{H}^{|V| \times d}, d \ll|V|$ in a unsupervised way.
 
\subsection{Self-guided Random Walk}
Before we propose the self-guided random walk, we first revisit the meta-path guided random walk. Meta-paths are used in heterogeneous graph embeddings since random walks on such graphs are biased to highly visible types of nodes. This means that sequences generated by random walk have a skewed distribution of nodes, with highly visible node types being over-represented. Node embeddings learned from these sequences will also be biased towards the highly visible domains. Thus, meta-paths are used to guide random walks to overcome this problem \cite{JUST}.

In this section, we propose a very simple but useful self-guided random walk method, which can adaptively change the probability of each node type in the next walk step and automatically balance the distribution of domains in the context.

A  self-guided walk on $G$ is a sequence of vertices $\left\langle v_{1}, v_{2}, \cdots, v_{k}\right\rangle$ such that $\left\langle v_{i}, v_{i+1}\right\rangle \in$ $E$ and node type $\phi(v_{i})$ appears $\mathcal{N}_{\phi(v_{i})}$ times in the current sequence, for $1 \leq i<k$. Instead of walking to all the neighbor nodes with equal probability like standard random walks, we define a modified probability for type $\phi$ nodes:
\begin{equation}
\displaystyle{\frac{e^{-\mathcal{N}_{\phi}}}{\sum\limits_{{\phi^{\prime}} \in T_{V}}e^{-\mathcal{N}_{\phi^{\prime}}}}},
\end{equation}

Assume we have $\left\langle v_{1}, v_{2}, \cdots, v_{k}\right\rangle$, after fraction reduction, the transition probability at step k+1 can be normalized to:
\begin{equation}
 p\left(v_{k+1} |\left\langle v_{1}, v_{2}, \cdots, v_{k}\right\rangle\right) =
\displaystyle{\frac{\frac{e^{-\mathcal{N}_{\phi_{v_{k+1}}}}}{|\{\phi_{v_{k+1}}=\phi_{v_{i}}| \left(v_{k},v_{i}\right) \in E\}|}}{\sum\limits_{ \left(v_{k},v_{j}\right) \in E}\frac{e^{-\mathcal{N}_{\phi_{v_{j}}}}}{{|\{\phi_{v_{j}}=\phi_{v_{i}}| \left(v_{k},v_{i}\right) \in E\}|}}}    },
\end{equation}

In this way, all the types of nodes will be balanced in the context. The prior work closest to this idea is JUST \cite{JUST}. They proposed a complex Jump \& Stay to balance the heterogeneous edges (connecting different types of nodes) and homogeneous edges (connecting the same type of nodes). However, our method focuses on achieving a balanced distribution of each type of node, which means that homo/heterogeneous edges will be automatically balanced.

In the next section, we first briefly introduce the \emph{hyperboloid} model and the gradient on it, then we explain how to train embedding from the results of the self-guided random walk.

\begin{table*}[htbp]
\centering
\small
\setlength{\tabcolsep}{1mm}
\begin{tabular}{ccccccc|cccccc}
\Xhline{2\arrayrulewidth}
Dataset      & \multicolumn{12}{c}{DBLP}                                                                                \\ \Xhline{2\arrayrulewidth}
Edge         & \multicolumn{6}{c|}{P-A}                            & \multicolumn{6}{c}{P-V}                            \\ \Xhline{2\arrayrulewidth}
Dimension    & 2      & 5      & 10     & 15     & 20     & 25     & 2      & 5      & 10     & 15     & 20     & 25     \\ 
Deepwalk     & 0.6933 & 0.8034 & 0.9324 & 0.9666 & 0.9722 & 0.9794 & 0.7324 & 0.7906 & 0.8813 & 0.9353 & 0.9505 & 0.9558 \\
LINE(1st)    & 0.5286 & 0.5397 & 0.6740 & 0.7220 & 0.7457 & 0.7668 & 0.5182 & 0.5500 & 0.7070 & 0.7295 & 0.7369 & 0.7436 \\
LINE(2nd)    & 0.6740 & 0.7379 & 0.7541 & 0.7868 & 0.7600 & 0.7621 & 0.6242 & 0.6349 & 0.6333 & 0.6343 & 0.6444 & 0.6440 \\
node2vec     & 0.7107 & 0.8162 & 0.9418 & 0.9719 & 0.9809 & 0.9881 & 0.7595 & 0.8019 & 0.8922 & 0.9382 & 0.9524 & 0.9596 \\
metapath2vec & 0.6686 & 0.8261 & 0.9202 & 0.9500 & 0.9623 & 0.9690 & 0.7286 & 0.9072 & 0.9691 & 0.9840 & 0.9879 & 0.9899 \\
PoincaréEmb  & 0.8251 & 0.8769 & 0.8921 & 0.8989 & 0.9024 & 0.9034 & 0.5718 & 0.5529 & 0.6271 & 0.6446 & 0.6600 & 0.6760 \\
HHNE         & \textbf{0.9835} & 0.9838 & 0.9887 & 0.9898 & 0.9913 & 0.9930 & 0.8449 & 0.9984 & 0.9985 & 0.9985 & 0.9985 & 0.9985 \\ 
JUST     & 0.7373 & 0.8682 & 0.9416 & 0.9664 & 0.9793 & 0.9889 & 0.6553 & 0.8007 & 0.8828 & 0.9265 & 0.9545 & 0.9567 \\ 
PTE & 0.5857 & 0.6702 & 0.7266 & 0.7483 & 0.7605 & 0.7590 & 0.6487 & 0.6758 & 0.6817 & 0.6905 & 0.7024 & 0.7135 \\ 
Hin2Vec & 0.7095 & 0.8465 & 0.9303 & 0.9561 & 0.9758 & 0.9748 & 0.6983 & 0.8435 & 0.9574 & 0.9760 & 0.9715 & 0.9802 \\ 
HeteSpaceyWalk & 0.6487 & 0.8128 & 0.9113 & 0.9459 & 0.9569 & 0.9543 & 0.7090 & 0.8826 & 0.9404 & 0.9725 & 0.9807 & 0.9835 \\ 
\textbf{Our method}          & 0.9602 & \textbf{0.9893} & \textbf{0.9972} & \textbf{0.9982} & \textbf{0.9985} & \textbf{0.9986} & \textbf{0.9728} & \textbf{0.9995} & \textbf{0.9999} & \textbf{0.9999} & \textbf{0.9999} & \textbf{0.9999} \\ \Xhline{2\arrayrulewidth}
\end{tabular}
\caption{AUC scores for the network reconstruction task on the DBLP network.}
\label{nrdblp}
\end{table*}

\subsection{Hyperbolic Embedding Learning}
Hyperbolic space has a negative curvature. There are multiple models that can be used to represent hyperbolic space, each having different advantages. The \emph{Poincar\'e ball} model is the best model for low dimensional visualizations of the embeddings. The \emph{Klein} model is often used for the calculation of Einstein midpoints because of its computational efficiency. The \emph{hyperboloid} model gives a closed-form of the gradient descent formula. Therefore, in this paper, we use the \emph{hyperboloid} model for the calculation of gradient. Below we go over several definitions needed to define the \emph{hyperboloid} model and the gradient on it \cite{gradient}.

\textbf{Definition 1.} The Minkowski inner product is defined as
\begin{equation}
\left<x,y\right>_{n:1} = x_1y_1\,+\,x_2y_2\,+\,\cdots\,+\,x_{n}y_{n}\,-\,x_{n+1}y_{n+1}.
\label{eq1}
\end{equation}
The Minkowski inner product is not a traditional inner product since it's not positive definite. 

\textbf{Definition 2.} The \emph{Hyperboloid} model is given by
\begin{equation}
\mathbb{\mathbb{H}}^n=\left\{x\in \mathbb{R}^{n:1}\,|\,\left<x,x\right>_{n:1} = -1,\,x_{n+1}>0 \right\}.
\end{equation}
$\mathbb{R}^{n:1}$ stands for the ambient Minkowski space. A simple \emph{hyperboloid} model is the hyperboloid of revolution: $\mathbb{\mathbb{H}}^2=\left\{x\in \mathbb{R}^3\,|\,x_1^2+x_2^2-x_3^2 = -1,\,x_3>0 \right\}.$

\textbf{Definition 3.} The hyperbolic distance between points $x$ and $y$ is defined as
\begin{equation}
d_{\mathrm{H}^{n}}(x,y)={\rm{cosh^{-1}}} \left(-\left<x,y\right>_{n:1}\right).
\end{equation}

We can calculate the gradient of distance function in $\mathbb{R}^{n:1}$:
\begin{align}
    \nabla_x^{\mathbb{R}^{n:1}}{d(x,y)}&=\nabla_x^{\mathbb{R}^{n:1}}{\rm{cosh^{-1}}} \left(-\left<x,y\right>_{n:1}\right)\nonumber\\
    &=-\frac 1 {\sqrt{\left<x,y\right>_{n:1}^2-1}} \nabla_x^{\mathbb{R}^{n:1}} \left<x,y\right>_{n:1}\nonumber\\
    &=-\frac y {\sqrt{\left<x,y\right>_{n:1}^2-1}}.\nonumber
\end{align}

\textbf{Definition 4.} The tangent space $T_x \mathbb{\mathbb{H}}^n$ to point $x\in \mathbb{\mathbb{H}}^n$ is the set of points satisfying
\begin{equation}
T_x\mathbb{\mathbb{H}}^n=\left\{u\in \mathbb{R}^{n:1}\,|\,\left<u,x\right>_{n:1} = 0\right\}.
\end{equation}
The projection from ambient space $\mathbb{R}^{n:1}$ to tangent space $T_x\mathbb{\mathbb{H}}^n$ is
\begin{equation}
\Pi_x(u)=u+\left<u,x\right>_{n:1}x.
\label{7}
\end{equation}

\textbf{Definition 5.} Exponential map is a common way to map a vector from tangent space to the \emph{hyperboloid} manifold. Considering $u \in T_x\mathbb{\mathbb{H}}^n$, it is exponential mapped as

\begin{equation}
{\rm cosh}(\sqrt{\left<u,u\right>_{n:1}})x+{\rm sinh}(\sqrt{\left<u,u\right>_{n:1}})\frac{u}{\sqrt{\left<u,u\right>_{n:1}}}.
\label{8}
\end{equation}
In this way, when we calculate the gradient of a given function $f$, first we can calculate its gradient in the ambient space $\nabla_x^{\mathbb{R}^{n:1}}f$. In the following steps, we need to project it to the tangent space using formula \ref{7} and apply exponential map on the vector we get using formula \ref{8}.

Once we are able to perform hyperbolic gradient descent, the next step is to design the loss function, which is the evaluation of an embedding. Following HEAT \cite{HEAT}, after generating the self-guided random walk sequences, a sliding window is used to examine all the sequences and add pairs of nodes that appear within the window to a multi-set $\mathcal{P}$ to serve as all the positive sample pairs for training. Intuitively, the further the distance between two nodes, the smaller the probability that there is a connection between them. Here we adopt that the probability should be proportional to $e^{-d_{\mathrm{H}^{n}}^{2}(\mathbf{e}_u,\mathbf{e}_{v})}$, where $\mathbf{e}_u$ and $\mathbf{e}_{v} $ are the embeddings of two nodes $u$ and $v$.
We can also easily deal with large networks using negative sampling. In this way, our loss function $\mathcal{L}$ is given by:
\begin{equation}
 \mathcal{L}=-\frac{1}{|\mathcal{P}|} \sum_{(u, v) \in \mathcal{P}} \log \frac{e^{-d_{\mathrm{H}^{n}}^{2}\left(\mathbf{e}_{u}, \mathbf{e}_{v}\right)  }}{\sum\limits_{v^{\prime} \in \mathcal{M}(u)} e^{ -d_{\mathrm{H}^{n}}^{2}\left(\mathbf{e}_{u}, \mathbf{e}_{{v^{\prime}}}\right)}}.
 \label{9}
\end{equation}
In formula \ref{9},   for a given $v$, $\mathcal{M}(u):=\{v\} \cup \{w|(u,w)\notin \mathcal{P}\}$ contains two parts. One is $v$, the other is the negative sampling set. The probability for a node $w$ to be chosen into the negative sample set $\mathcal{M}(u)$ is proportional to the frequency of that node $w$ in the full sample set $\mathcal{P}$. Since we already have the gradient of distance function in $\mathbb{R}^{n:1}$ and the methodology to project it to the \emph{hyperboloid}, the gradient of the loss function can be easily calculated by chain rule.

\section{Experiments}
In this section, we evaluate the performance of our proposed method both qualitatively through visualization, and quantitatively on the tasks of network reconstruction and link prediction. We compare our method with several baselines and explore the sensitivity of our method to the choice of parameters. For a fair comparison, we use the exact same datasets, experiments, settings, and metrics as HHNE.

\subsection{Datasets}

The following datasets were given to us by the authors' of HHNE:
\begin{itemize}
    \item \textbf{DBLP}: This is a subset network of DBLP, which contains three node types: 14,475 authors (A), 14,376 papers (P), and 20 venues (V). The network also contains the following edge types: 41,794 \emph{paper authorship} relations (P-A) and 14,376 \emph{publish} relations (P-V). 
    \item \textbf{MovieLens}: This is a subset network of MovieLens, which contains three node types: 11,718 actors (A), 9,160 movies (M), and 3,510 directors (D). The network also contains the following edge types: 64,051  \emph{act in} relations (M-A) and 9,160 \emph{direct} relations (M-D).
\end{itemize}

\begin{table*}[]
\centering
\small
\setlength{\tabcolsep}{1mm}
\begin{tabular}{ccccccc|cccccc}
\Xhline{2\arrayrulewidth}
Dataset      & \multicolumn{12}{c}{MoiveLens}                                                                           \\ \Xhline{2\arrayrulewidth}
Edge         & \multicolumn{6}{c|}{M-A}                            & \multicolumn{6}{c}{M-D}                            \\ \Xhline{2\arrayrulewidth}
Dimension    & 2      & 5      & 10     & 15     & 20     & 25     & 2      & 5      & 10     & 15     & 20     & 25     \\ 
Deepwalk     & 0.6320 & 0.6763 & 0.7610 & 0.8244 & 0.8666 & 0.8963 & 0.6626 & 0.7263 & 0.8246 & 0.8784 & 0.9117 & 0.9345 \\
LINE(1st)    & 0.5424 & 0.5675 & 0.6202 & 0.6593 & 0.6925 & 0.7251 & 0.5386 & 0.5839 & 0.6114 & 0.6421 & 0.6748 & 0.7012 \\
LINE(2nd)    & 0.6378 & 0.7047 & 0.7739 & 0.7955 & 0.8065 & 0.8123 & 0.6016 & 0.6521 & 0.6969 & 0.7112 & 0.7503 & 0.7642 \\
node2vec     & 0.6402 & 0.6774 & 0.7653 & 0.8304 & 0.8742 & 0.9035 & 0.6707 & 0.7283 & 0.8308 & 0.8867 & 0.9186 & 0.9402 \\
metapath2vec & 0.6404 & 0.6578 & 0.7231 & 0.7793 & 0.8189 & 0.8483 & 0.6589 & 0.7230 & 0.8063 & 0.8455 & 0.8656 & 0.8800 \\
PoincaréEmb  & 0.5231 & 0.5317 & 0.5404 & 0.5479 & 0.5522 & 0.5545 & 0.6213 & 0.7266 & 0.7397 & 0.7378 & 0.7423 & 0.7437 \\
HHNE         & 0.8832 & 0.9168 & 0.9211 & 0.9221 & 0.9239 & 0.9233 & \textbf{0.9952} & 0.9968 & 0.9975 & 0.9972 & 0.9982 & 0.9992 \\ 
JUST     & 0.7108 & 0.8451 & 0.9213 & 0.9514 & 0.9790 & 0.9832 & 0.7267 & 0.8538 & 0.9106 & 0.9510 & 0.9735 & 0.9874 \\ 
PTE      & 0.5601 & 0.6078 & 0.6496 & 0.6768 & 0.7023 & 0.7127 & 0.6035 & 0.6540 & 0.6830 & 0.7149 & 0.7388 & 0.7404 \\ 
Hin2Vec  & 0.7345 & 0.8628 & 0.9204 & 0.9571 & 0.9677 & 0.9702 & 0.7440 & 0.8658 & 0.9127 & 0.9438 & 0.9615 & 0.9735 \\ 
HeteSpaceyWalk & 0.6556 & 0.7121 & 0.7888 & 0.8331 & 0.8688 & 0.8997 & 0.6830 & 0.7568 & 0.8399 & 0.8817 & 0.9005 & 0.9138 \\   
\textbf{Our method}          & \textbf{0.9251} & \textbf{0.9690} & \textbf{0.9873} & \textbf{0.9934} & \textbf{0.9959} & \textbf{0.9970} & 0.9904 & \textbf{0.9996} & \textbf{0.9999} & \textbf{0.9999} & \textbf{0.9999} & \textbf{0.9999} \\ \Xhline{2\arrayrulewidth}

\end{tabular}
\caption{AUC scores for the network reconstruction task on the MovieLens network.}
\label{nrmovie}
\end{table*}

\begin{table*}[]
\centering
\small
\setlength{\tabcolsep}{1mm}
\begin{tabular}{ccccccc|cccccc}
\Xhline{2\arrayrulewidth}
Dataset      & \multicolumn{12}{c}{DBLP}                                                                                \\ \Xhline{2\arrayrulewidth}
Edge         & \multicolumn{6}{c|}{P-A}                            & \multicolumn{6}{c}{P-V}                            \\ \Xhline{2\arrayrulewidth}
Dimension    & 2      & 5      & 10     & 15     & 20     & 25     & 2      & 5      & 10     & 15     & 20     & 25     \\
Deepwalk     & 0.5813 & 0.7370 & 0.8250 & 0.8664 & 0.8807 & 0.8878 & 0.7075 & 0.7197 & 0.7292 & 0.7325 & 0.7522 & 0.7640 \\
LINE(1st)    & 0.5090 & 0.5168 & 0.5427 & 0.5631 & 0.5742 & 0.5857 & 0.5160 & 0.5663 & 0.5873 & 0.5896 & 0.5891 & 0.5846 \\
LINE(2nd)    & 0.5909 & 0.6351 & 0.6510 & 0.6582 & 0.6644 & 0.6782 & 0.5121 & 0.5216 & 0.5332 & 0.5425 & 0.5492 & 0.5512 \\
node2vec     & 0.6709 & 0.7527 & 0.8469 & 0.8881 & 0.9037 & 0.9102 & 0.7369 & 0.7286 & 0.7481 & 0.7583 & 0.7674 & 0.7758 \\
metapath2vec & 0.6536 & 0.7294 & 0.8279 & 0.8606 & 0.8740 & 0.8803 & 0.7059 & 0.8516 & 0.9248 & 0.9414 & 0.9504 & 0.9536 \\
PoincaréEmb  & 0.6742 & 0.7381 & 0.7699 & 0.7743 & 0.7806 & 0.7830 & 0.8257 & 0.8878 & 0.9113 & 0.9142 & 0.9185 & 0.9192 \\
HHNE         & 0.8777 & 0.9041 & 0.9111 & 0.9111 & 0.9106 & 0.9117 & 0.9331 & 0.9409 & 0.9619 & 0.9625 & 0.9620 & 0.9612 \\ 
JUST     & 0.6577 & 0.7538 & 0.7926 & 0.8041 & 0.8056 & 0.8026 & 0.5847 & 0.7060 & 0.7468 & 0.7543 & 0.7610 & 0.7628 \\ 
PTE & 0.5070 & 0.5369 & 0.5509 & 0.5684 & 0.5869 & 0.5816 & 0.5425 & 0.6209 & 0.6407 & 0.6584 & 0.6693 & 0.6761 \\ 
Hin2Vec & 0.6412 & 0.7690 & 0.8187 & 0.8269 & 0.8315 & 0.8341 & 0.6544 & 0.6800 & 0.7399 & 0.7230 & 0.7405 & 0.7558 \\ 
HeteSpaceyWalk & 0.6337 & 0.7573 & 0.8302 & 0.8792 & 0.8916 & 0.9027 & 0.6931 & 0.8474 & 0.9042 & 0.9201 & 0.9344 & 0.9450 \\  
\textbf{Our method}          & \textbf{0.8803} & \textbf{0.9129} & \textbf{0.9264} & \textbf{0.9290} & \textbf{0.9294} & \textbf{0.9303} & \textbf{0.9547} & \textbf{0.9624} & \textbf{0.9681} & \textbf{0.9701} & \textbf{0.9742} & \textbf{0.9751} \\ \Xhline{2\arrayrulewidth}
\end{tabular}
\caption{AUC scores for the link prediction task on the DBLP network.}
\label{lpdblp}
\end{table*}

\begin{table*}[]
\centering
\small
\setlength{\tabcolsep}{1mm}
\begin{tabular}{ccccccc|cccccc}
\Xhline{2\arrayrulewidth}
Dataset      & \multicolumn{12}{c}{MoiveLens}                                                                           \\ \Xhline{2\arrayrulewidth}
Edge         & \multicolumn{6}{c|}{M-A}                            & \multicolumn{6}{c}{M-D}                            \\ \Xhline{2\arrayrulewidth}
Dimension    & 2      & 5      & 10     & 15     & 20     & 25     & 2      & 5      & 10     & 15     & 20     & 25     \\ 
Deepwalk     & 0.6278 & 0.6353 & 0.6680 & 0.6791 & 0.6868 & 0.6890 & 0.6258 & 0.6482 & 0.6976 & 0.7163 & 0.7324 & 0.7446 \\
LINE(1st)    & 0.5053 & 0.5636 & 0.5914 & 0.6184 & 0.6202 & 0.6256 & 0.5139 & 0.5496 & 0.5885 & 0.6647 & 0.6742 & 0.6957 \\
LINE(2nd)    & 0.5712 & 0.5874 & 0.6361 & 0.6442 & 0.6596 & 0.6700 & 0.6501 & 0.6607 & 0.7499 & 0.7756 & 0.7982 & 0.8051 \\
node2vec     & 0.6349 & 0.6402 & 0.6700 & 0.6814 & 0.6910 & 0.6977 & 0.6299 & 0.6589 & 0.7034 & 0.7241 & 0.7412 & 0.7523 \\
metapath2vec & 0.6168 & 0.6212 & 0.6332 & 0.6382 & 0.6453 & 0.6508 & 0.6191 & 0.6332 & 0.6687 & 0.6702 & 0.6746 & 0.6712 \\
PoincaréEmb  & 0.5535 & 0.5779 & 0.5984 & 0.5916 & 0.5988 & 0.5995 & 0.5856 & 0.6290 & 0.6518 & 0.6715 & 0.6821 & 0.6864 \\
HHNE         & 0.7715 & 0.8255 & 0.8312 & 0.8319 & 0.8318 & 0.8309 & 0.8520 & 0.8967 & 0.8984 & 0.9007 & 0.9000 & 0.9018 \\ 
JUST     & 0.6297 & 0.7199 & 0.8004 & 0.7948 & 0.8057 & 0.8089 & 0.6217 & 0.7562 & 0.8320 & 0.8361 & 0.8394 & 0.8403 \\ 
PTE & 0.5035 & 0.5100 & 0.5428 & 0.5462 & 0.5436 & 0.5542 & 0.5529 & 0.6450 & 0.6426 & 0.6480 & 0.6477 & 0.6529 \\ 
Hin2Vec & 0.6534 & 0.7486 & 0.7666 & 0.7693 & 0.7614 & 0.7784 & 0.6792 & 0.7645 & 0.8030 & 0.8081 & 0.8002 & 0.8179 \\ 
HeteSpaceyWalk & 0.6135 & 0.6247 & 0.6221 & 0.6374 & 0.6429 & 0.6433 & 0.6188 & 0.6296 & 0.6536 & 0.6614 & 0.6709 & 0.6754 \\  
\textbf{Our method}       & \textbf{0.8498} & \textbf{0.8707} & \textbf{0.8720} & \textbf{0.8723} & \textbf{0.8747}& \textbf{0.8765} & \textbf{0.8781} & \textbf{0.9023} & \textbf{0.9058} & \textbf{0.9083} & \textbf{0.9096} & \textbf{0.9124} \\ \Xhline{2\arrayrulewidth}
\end{tabular}
\caption{AUC scores for the link prediction task on the MovieLens network.}
\label{lpmoive}
\end{table*}

\subsection{Experiment Settings}

We compared our method with several homogeneous and heterogeneous network embedding methods. The homogeneous methods include DeepWalk, LINE, node2vec, and the hyperbolic embedding method, PoincaréEmb. The heterogeneous methods include metapath2vec, JUST, PTE, Hin2Vec, HeteSpaceyWalk, and the hyperbolic embedding method HHNE. Note that many recent works on Graph Neural Networks focusing on designing a learning mechanism on graphs, in most cases they are end-to-end, supervised or semi-supervised. Thus we do not include them in baselines as we are interested in comparing the performance of our method against other network embedding methods 
that are unsupervised and focusing on transforming networks into low-dimensional space.

For methods based on meta-path guided random walks, we use ``APA" for the relation ``P-A", and ``APVPA" for the relation ``P-V" in network reconstruction and link prediction experiments on the DBLP dataset. For the experiments on the MovieLens dataset, we use ``AMDMA" for both relations ``M-A" and ``M-D". (Same as Wang et al. (\citeyear{HHNE})). For PTE, we use the unsupervised setting and construct two bipartite graphs for each dataset: DBLP (A-P, P-V), MovieLens (A-M, M-D). 

The performance of DeepWalk, LINE, node2vec, metapath2vec, PoincaréEmb, and HHNE reported here is taken from  Wang et al. (\citeyear{HHNE}) (since we used the same datasets and settings). For JUST, PTE, Hin2Vec, and HeteSpaceyWalk, we tried our best to fine-tune the parameters and report the best results. For our method, we use the following parameters:  We do 10 random walks from each node in the training set with the length of 80, and use a sliding window of size 5 to generate positive samples. For the \emph{hyperboloid} embedding learning, we generate 20 negative samples for each positive one, and use the learning rate of 0.3 and a batch size of 512 to train 5 epochs.
All the experiments are run on an Amazon AWS ``p2.8xlarge" instance running a Linux OS with 488GB of RAM, and the random seeds are set to 0 at the beginning.

\subsection{Network Reconstruction}
The network reconstruction task tests the ability of embedding methods to preserve the original network structure. That is, to recover the structure of a network from embeddings learned on the whole network. To test our method, we use all the edges between two node types as the positive set and all the non-edges as the negative set. We use the distances between node embeddings (on the \emph{hyperboloid}) to get an AUC score by thresholding on different distance values for predicting whether a link exists between two nodes.

The results for the DBLP and MoiveLens networks are shown in Table \ref{nrdblp} and Table \ref{nrmovie}, respectively. Our models outperform all the baselines for all embedding dimensions from 2 to 25, with two exceptions (out of 12) where HHNE generates better results when embedding in 2 dimensions. Note that even though HHNE requires task-specific meta-paths, our model outperforms it in most cases without any external knowledge. Our better results may be attributed to the accurate gradient calculation in the \emph{hyperboloid} and the more diverse context generated by the self-guided random walk. HHNE also outperforms other baselines in most cases, suggesting the superiority of hyperbolic heterogeneous network embedding methods in low dimensions. Another hyperbolic embedding method, PoincaréEmb, performs much worse than HHNE and our method. The reason for this may be that this method is designed for homogeneous networks and only preserves the first-order proximity, suggesting that to embed heterogeneous networks into hyperbolic space, either meta-paths or our proposed self-guided random walk should be utilized. Also, other models tend to have better performance than LINE, PTE, and PoincaréEmb, because these three models only capture first or second-order proximities, while others capture longer context information between nodes.
\subsection{Link Prediction} 
The link prediction task tests the ability of embedding methods to predict unseen network structure. For each type of relation, we randomly remove 20\% of the edges from the network without increasing the number of connected components, the network is then used for learning embeddings. For testing, all the removed edges are used as the positive set and a negative set is created by randomly sampling an equal number of non-edges. The AUC score is calculated in the same manner as for the network reconstruction task.

The results for the DBLP and MoiveLens networks for this task are shown in Table \ref{lpdblp} and Table \ref{lpmoive}, respectively. Overall, the results of link prediction experiments are consistent with the results of network reconstruction experiments. Our models outperform all the baselines, with the greatest difference seen in lower dimensions. As with the prior task, HHNE ranked second across all the methods. LINE, PTE, and PoincaréEmb, which only capture 1-hop or 2-hop neighbourhood of nodes perform worse than other baselines. Link prediction is in general a harder task than network reconstruction, since it involves the prediction of unseen edges. Thus, we see that for each method under the same edge type and dimensions, the AUC score for link prediction is almost always lower than the AUC of network reconstruction.

\subsection{Visualization}
\begin{figure}[htbp]

\centerline{\includegraphics[width=0.8\linewidth]{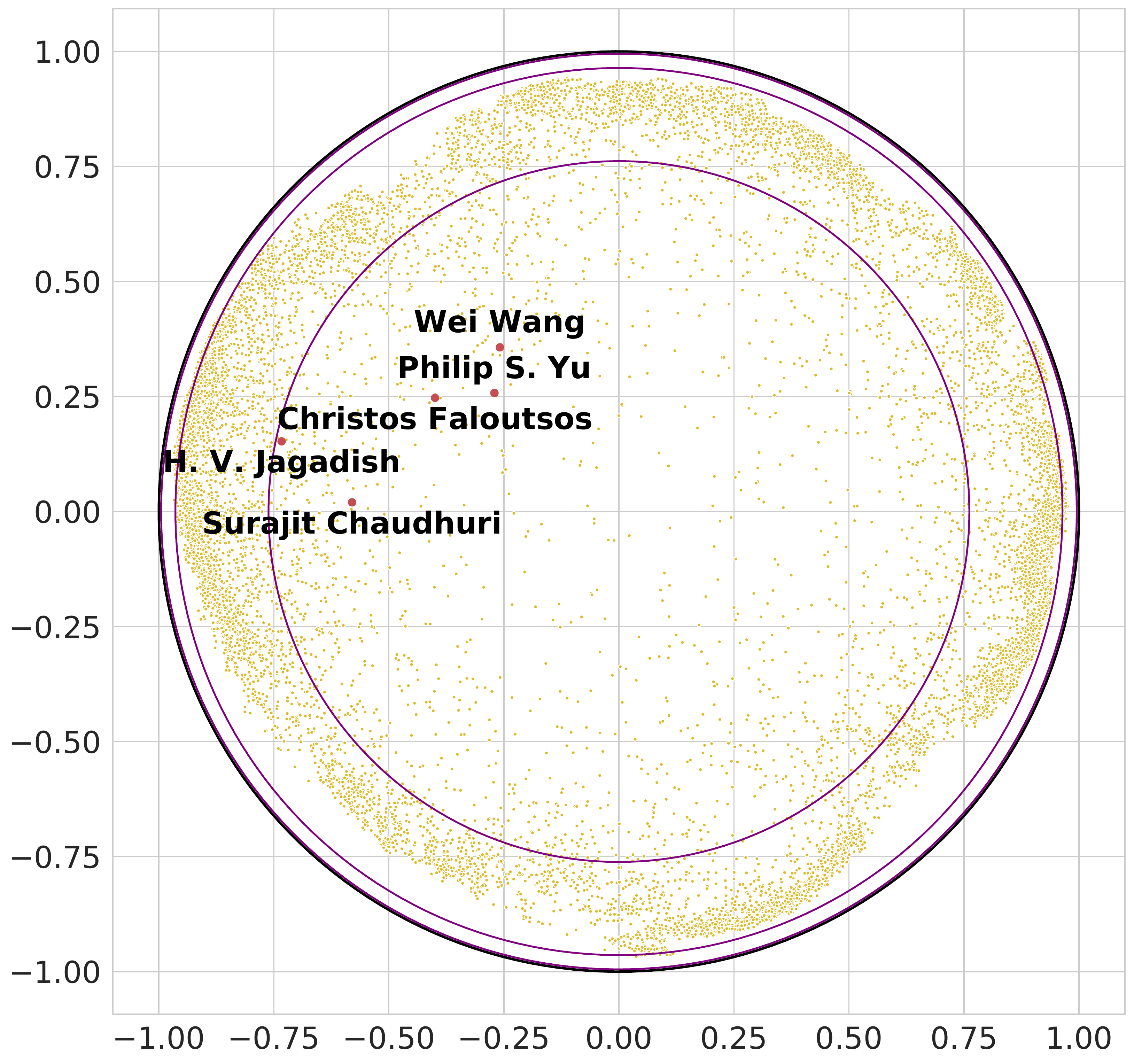}} 

\caption{Embedding of each author in DBLP after projected to a 2-dimensional \emph{Poincar\'e disk}, each dot represent an author}
\label{embedding}
\end{figure}

\begin{figure}[htbp]

\centerline{\includegraphics[width=0.8\linewidth]{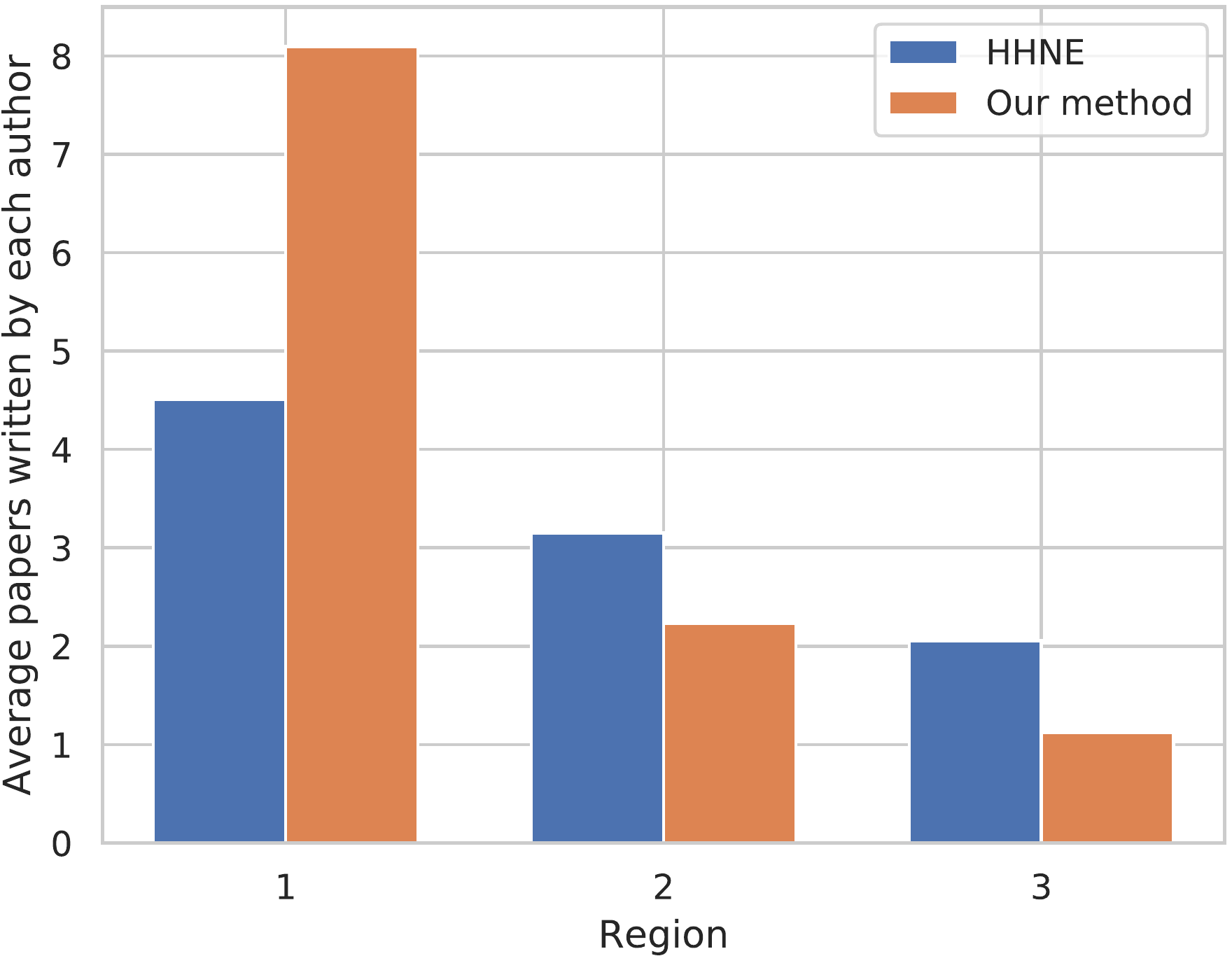}} 

\caption{The results of HHNE and our methods for average number of papers written by each author}
\label{bar}
\end{figure}

\begin{figure*}[htbp]
\centering

\includegraphics[width=0.95\textwidth]{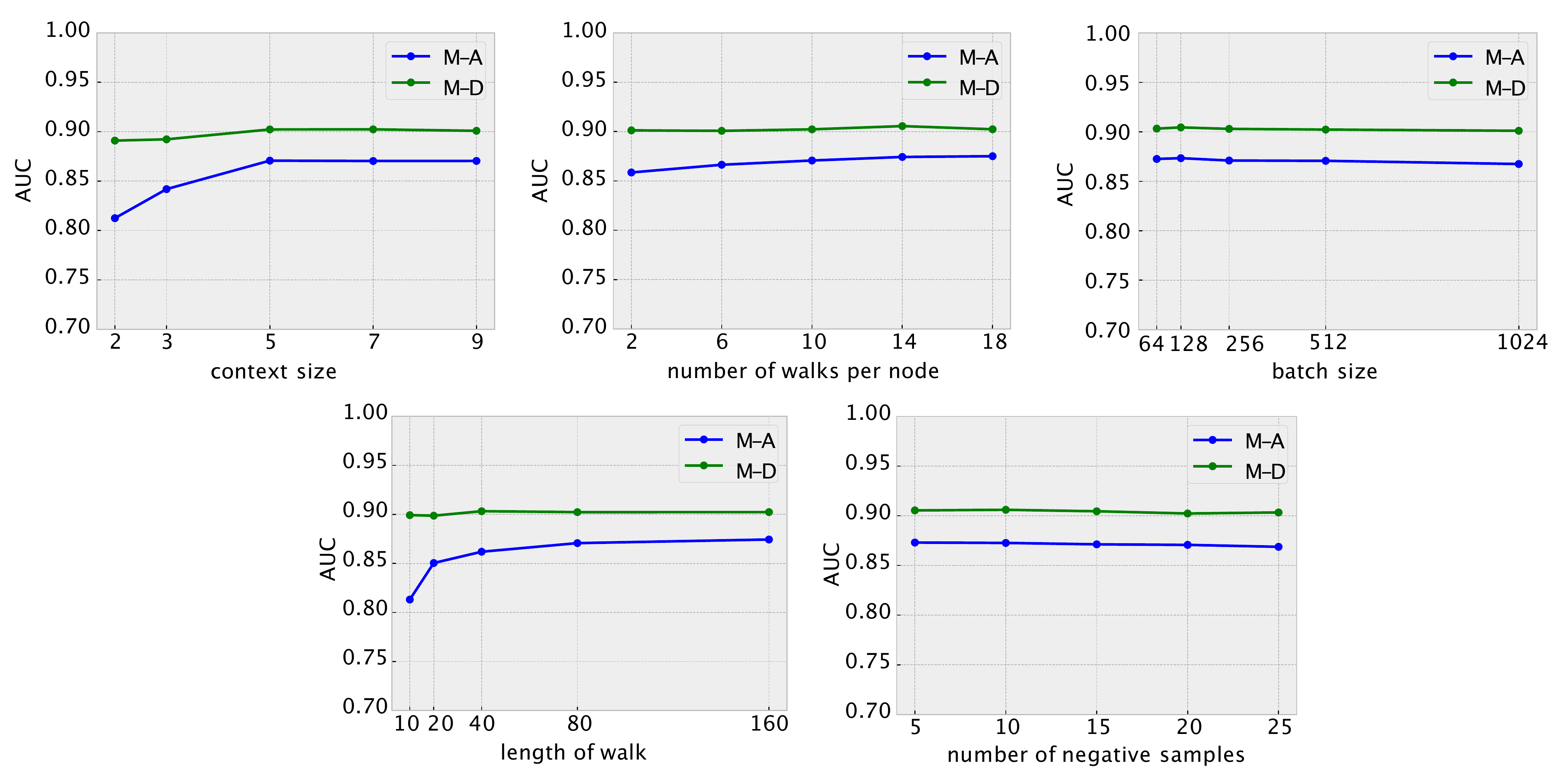} 

\centering
\caption{Parameter sensitivity for the link prediction task on the MovieLens dataset.}
\label{para}
\end{figure*}

We repeat the visualization experiment from HHNE to evaluate the latent hierarchy learned by our model. We show our model's embedding of each author in the DBLP network projected to a 2-dimensional \emph{Poincar\'e disk} in Figure \ref{embedding}. We divide the disk into three equal-radius regions using the distance function on a \emph{Poincar\'e disk}, shown in Equation \ref{disball} below:
\begin{equation}d_{\mathbb{H}}\left(u, v\right)=\cosh ^{-1}\left(1+2 \frac{\left\|u-v\right\|^{2}}{\left(1-\left\|u\right\|^{2}\right)\left(1-\left\|v\right\|^{2}\right)}\right)  \label{disball} \end{equation}

\noindent \textbf{Region 1}: $d_{\mathbb{H}}\left(u, 0\right) \leq 2 $. \textbf{Region 2}: $ 2 \leq d_{\mathbb{H}}\left(u, 0\right) \leq 4 $. \textbf{Region 3}: $ 4 \leq d_{\mathbb{H}}\left(u, 0\right) \leq 6 $

We mark the three boundaries in Figure \ref{embedding} using purple circles\footnote{Note that regions look disproportionate because they are visualized on a 2D Euclidean grid. Also note that the hyperbolic space spans exponentially, so even though the embeddings in Region 3 look overlapped, they are not.}. The latent hierarchy of authors is based on their degree (i.e., number of publications). The distance between the authors and the origin of the disk reflects the latent hierarchy of the authors captured by our model. The closer an author to the origin, the higher its latent hierarchy. This means that authors in Region 1 have higher latent hierarchy than authors in Region 2 and so on. In Region 1, we can find all the five high-impact authors mentioned by HHNE (labelled in Figure \ref{embedding}). Furthermore, we calculate the average number of papers written by the authors in each of the three regions (1,804, 11,704, and 967 authors respectively) and show that in Figure \ref{bar}. We also show the same measurement for HHNE from Wang et al. (\citeyear{HHNE}). The average number of papers written by the authors in Region 1 is greater for our model compared to HHNE. Also, as seen in Figure \ref{bar}, our model better captures the power-law degree distribution inherent in authorship networks. This means that the latent hierarchy learned by our model is better than the one learned by HHNE.

\subsection{Parameter Sensitivity Analysis}

Our model has several hyper-parameters:

\noindent\textbf{batch size -} The batch size used for training.

\noindent\textbf{context size -} The window size used for collecting the positive samples.

\noindent\textbf{number of walks per node -} The number of random walks started at each node.

\noindent\textbf{length of walk -} The length of each random walk.

\noindent\textbf{negative samples -} The number of negative samples per positive sample.

\smallskip
\noindent The parameter sensitivity analysis for our model for the link prediction task on the MovieLens dataset (for M-A and M-D edges) is shown in Fig. \ref{para}. We tune each parameter separately while fixing the other parameters to their default values and fix the dimension of embeddings to 5. Overall, our model is not parameter-sensitive. The most sensitive parameters are the context size and the length of the random walk. These two parameters need to be set to at least 5 and 10 (respectively) to get stable results. Generally, a larger context enables a better representation of neighbourhood information, and a longer length of walk enables the self-guided random walk to generate more balanced node sequences for each type of node. 

\section{Conclusion}

In this paper, we propose a novel hyperbolic embedding method for heterogeneous networks. Our model uses a variant of random walk called self-guided random walk. This novel random walk method removes the need for meta-paths, enabling us to get consistent embeddings that can be used for different down-stream tasks without the need for retraining our model. Lack of reliance on meta-paths also removes the need for external knowledge. Through thorough experimentation using publicly available datasets, we show that our model outperforms a variety of well-known baselines, including the only other hyperbolic heterogeneous network embedding method, HHNE. Future avenues for research could involve the extension of our framework to embed time-evolving heterogeneous networks into hyperbolic space. The code and data for this paper will be made available upon request.

 \section{Acknowledgements}
 We thank the authors of HHNE \cite{HHNE} for sharing their implementation/datasets and answering our questions about their paper. This research was supported in part by the Dartmouth Burke Research Initiation Award.
\bibliography{simple}
\end{document}